\documentclass[conference]{IEEEtran}
\IEEEoverridecommandlockouts

\usepackage[disable]{todonotes}
\newcommand{\alp}[2][] {\todo[inline,backgroundcolor=blue!20!white, #1]{(Alp) #2}}
\newcommand{\wouter}[2][] {\todo[inline,backgroundcolor=green!20!white, #1]{(Wouter) #2}}
\newcommand{\lense}[2][] {\todo[inline,backgroundcolor=red!20!white, #1]{(Lense) #2}}

\usepackage{cite}
\usepackage{amsmath,amssymb,amsfonts}
\usepackage{graphicx}
\usepackage{textcomp}
\usepackage{xcolor}
\usepackage{pgfplots}
\usepackage{placeins}
\usepackage{subcaption}  
\usepackage{adjustbox}
\usepackage{amsmath}
\usepackage{xcolor}
\usepackage{mathtools}
\usepackage{amssymb}
\usepackage{amsthm}
\usepackage{booktabs}
\usepackage{url}
\usepackage{algorithm}
\usepackage{algpseudocode}

\def\0{{\mathbf 0}}
\def\1{{\mathbf 1}}

\def\c{{\mathrm c}}

\def\F{{\mathcal F}}

\def\-{\text{-}}
\def\+{\text{+}}
\newcommand\given[1][]{\:#1\vert\:}

\pgfplotsset{compat=1.16}
\usetikzlibrary{intersections}
\usepackage{bm}
\usetikzlibrary{positioning}
\usepgfplotslibrary{fillbetween}
\usepgfplotslibrary{groupplots}
\usetikzlibrary{shapes.geometric,backgrounds}
\usepackage[outline]{contour}

\usepackage{tikz}
\usepackage{xifthen}

\pgfdeclarelayer{bg}    
\pgfsetlayers{bg,main}  
\definecolor{beige}{RGB}{245, 245, 220}

\definecolor{darkgrey}{RGB}{75, 75, 75}
\definecolor{lightgrey}{RGB}{250, 250, 250}

\usetikzlibrary{calc, arrows, fit, positioning, patterns, decorations.pathreplacing, shapes}
\tikzstyle{dash} = [dashed, -latex,>=latex]
\tikzstyle{line} = [draw, -latex,>=latex]
\tikzstyle{smallbox} = [draw, minimum size=5.0mm]
\tikzstyle{box} = [draw, minimum size=7.0mm]
\tikzstyle{bigbox} = [draw, minimum size=10.0mm]
\tikzstyle{switch} = [trapezium, trapezium angle=120, draw, rotate=90,  inner ysep=5pt, outer sep=5pt,
minimum height=7mm, minimum width=7mm]
\tikzstyle{roundbox} = [draw, circle, inner sep=0pt, minimum size=3mm]
\tikzstyle{clamped} = [draw, fill=black, minimum size=0.15cm]
\tikzstyle{msgcircle} = [shape=circle, draw, inner sep=0pt, minimum size=4mm, fill=white, font=\scriptsize]
\tikzstyle{darkmsgcircle} = [shape=circle, draw, inner sep=0pt, minimum size=4mm, fill=darkgrey, text=white, font=\scriptsize]
\tikzstyle{msgdoublecircle} = [shape=circle, double, double distance=1.5pt, draw, inner sep=0pt, minimum size=5mm, fill=white]
\tikzstyle{darkmsgdoublecircle} = [shape=circle, double, double distance=1.5pt, draw, inner sep=0pt, minimum size=5mm, fill=darkgrey, text=white, font=\bfseries]

\newcommand{\msg}[6]{
      \ifthenelse{\isin{#1}{left} \AND \isin{#2}{down}}{
            \coordinate (anchor) at ($({#3})!{#5}!({#4})$);
            \node[xshift=-6.0mm] at (anchor) {#6};
            \node[xshift=-1.0mm] at (anchor) {$\downarrow$};
      }{}
      \ifthenelse{\isin{#1}{right} \AND \isin{#2}{down}}{
            \coordinate (anchor) at ($({#3})!{#5}!({#4})$);
            \node[xshift=6.0mm] at (anchor) {#6};
            \node[xshift=1.0mm] at (anchor) {$\downarrow$};
      }{}

      \ifthenelse{\isin{#1}{down} \AND \isin{#2}{right}}{
            \coordinate (anchor) at ($({#3})!{#5}!({#4})$);
            \node[ yshift=-4.0mm] at (anchor) {#6};
            \node[yshift=-1.0mm] at (anchor) {$\rightarrow$};
      }{}
      \ifthenelse{\isin{#1}{up} \AND \isin{#2}{right}}{
            \coordinate (anchor) at ($({#3})!{#5}!({#4})$);
            \node[ yshift=4.0mm] at (anchor) {#6};
            \node[yshift=1.0mm] at (anchor) {$\rightarrow$};
      }{}

      \ifthenelse{\isin{#1}{down} \AND \isin{#2}{left}}{
            \coordinate (anchor) at ($({#3})!{#5}!({#4})$);
            \node[ yshift=-4.0mm] at (anchor) {#6};
            \node[yshift=-1.0mm] at (anchor) {$\leftarrow$};
      }{}
      \ifthenelse{\isin{#1}{up} \AND \isin{#2}{left}}{
            \coordinate (anchor) at ($({#3})!{#5}!({#4})$);
            \node[ yshift=4.0mm] at (anchor) {#6};
            \node[yshift=1.0mm] at (anchor) {$\leftarrow$};
      }{}

      \ifthenelse{\isin{#1}{left} \AND \isin{#2}{up}}{
            \coordinate (anchor) at ($({#3})!{#5}!({#4})$);
            \node[ xshift=-6.0mm] at (anchor) {#6};
            \node[xshift=-1.0mm] at (anchor) {$\uparrow$};
      }{}
      \ifthenelse{\isin{#1}{right} \AND \isin{#2}{up}}{
            \coordinate (anchor) at ($({#3})!{#5}!({#4})$);
            \node[ xshift=6.0mm] at (anchor) {#6};
            \node[xshift=1.0mm] at (anchor) {$\uparrow$};
      }{}
}

\newcommand{\msgcircle}[6]{
      \ifthenelse{\isin{#1}{left} \AND \isin{#2}{down}}{
            \coordinate (anchor) at ($({#3})!{#5}!({#4})$);
            \node[msgcircle,xshift=-5.0mm] at (anchor) {#6};
            \node[xshift=-1.5mm] at (anchor) {$\downarrow$};
      }{}
      \ifthenelse{\isin{#1}{right} \AND \isin{#2}{down}}{
            \coordinate (anchor) at ($({#3})!{#5}!({#4})$);
            \node[msgcircle,xshift=5.0mm] at (anchor) {#6};
            \node[xshift=1.5mm] at (anchor) {$\downarrow$};
      }{}

      \ifthenelse{\isin{#1}{down} \AND \isin{#2}{right}}{
            \coordinate (anchor) at ($({#3})!{#5}!({#4})$);
            \node[msgcircle, yshift=-5.0mm] at (anchor) {#6};
            \node[yshift=-2.0mm] at (anchor) {$\rightarrow$};
      }{}
      \ifthenelse{\isin{#1}{up} \AND \isin{#2}{right}}{
            \coordinate (anchor) at ($({#3})!{#5}!({#4})$);
            \node[msgcircle, yshift=5.0mm] at (anchor) {#6};
            \node[yshift=2.0mm] at (anchor) {$\rightarrow$};
      }{}

      \ifthenelse{\isin{#1}{down} \AND \isin{#2}{left}}{
            \coordinate (anchor) at ($({#3})!{#5}!({#4})$);
            \node[msgcircle, yshift=-5.0mm] at (anchor) {#6};
            \node[yshift=-2.0mm] at (anchor) {$\leftarrow$};
      }{}
      \ifthenelse{\isin{#1}{up} \AND \isin{#2}{left}}{
            \coordinate (anchor) at ($({#3})!{#5}!({#4})$);
            \node[msgcircle, yshift=5.0mm] at (anchor) {#6};
            \node[yshift=2.0mm] at (anchor) {$\leftarrow$};
      }{}

      \ifthenelse{\isin{#1}{left} \AND \isin{#2}{up}}{
            \coordinate (anchor) at ($({#3})!{#5}!({#4})$);
            \node[msgcircle, xshift=-5.0mm] at (anchor) {#6};
            \node[xshift=-1.5mm] at (anchor) {$\uparrow$};
      }{}
      \ifthenelse{\isin{#1}{right} \AND \isin{#2}{up}}{
            \coordinate (anchor) at ($({#3})!{#5}!({#4})$);
            \node[msgcircle, xshift=5.0mm] at (anchor) {#6};
            \node[xshift=1.5mm] at (anchor) {$\uparrow$};
      }{}
}

\newcommand{\darkmsg}[6]{
      \ifthenelse{\isin{#1}{left} \AND \isin{#2}{down}}{
            \coordinate (anchor) at ($({#3})!{#5}!({#4})$);
            \node[darkmsgcircle, xshift=-5mm] at (anchor) {#6};
            \node[xshift=-1.5mm] at (anchor) {$\downarrow$};
      }{}
      \ifthenelse{\isin{#1}{right} \AND \isin{#2}{down}}{
            \coordinate (anchor) at ($({#3})!{#5}!({#4})$);
            \node[darkmsgcircle, xshift=5mm] at (anchor) {#6};
            \node[xshift=1.5mm] at (anchor) {$\downarrow$};
      }{}

      \ifthenelse{\isin{#1}{down} \AND \isin{#2}{right}}{
            \coordinate (anchor) at ($({#3})!{#5}!({#4})$);
            \node[darkmsgcircle, yshift=-5.0mm] at (anchor) {#6};
            \node[yshift=-2.0mm] at (anchor) {$\rightarrow$};
      }{}
      \ifthenelse{\isin{#1}{up} \AND \isin{#2}{right}}{
            \coordinate (anchor) at ($({#3})!{#5}!({#4})$);
            \node[darkmsgcircle, yshift=5.0mm] at (anchor) {#6};
            \node[yshift=2.0mm] at (anchor) {$\rightarrow$};
      }{}

      \ifthenelse{\isin{#1}{down} \AND \isin{#2}{left}}{
            \coordinate (anchor) at ($({#3})!{#5}!({#4})$);
            \node[darkmsgcircle, yshift=-5.0mm] at (anchor) {#6};
            \node[yshift=-2.0mm] at (anchor) {$\leftarrow$};
      }{}
      \ifthenelse{\isin{#1}{up} \AND \isin{#2}{left}}{
            \coordinate (anchor) at ($({#3})!{#5}!({#4})$);
            \node[darkmsgcircle, yshift=5.0mm] at (anchor) {#6};
            \node[yshift=2.0mm] at (anchor) {$\leftarrow$};
      }{}

      \ifthenelse{\isin{#1}{left} \AND \isin{#2}{up}}{
            \coordinate (anchor) at ($({#3})!{#5}!({#4})$);
            \node[darkmsgcircle, xshift=-5.0mm] at (anchor) {#6};
            \node[xshift=-1.5mm] at (anchor) {$\uparrow$};
      }{}
      \ifthenelse{\isin{#1}{right} \AND \isin{#2}{up}}{
            \coordinate (anchor) at ($({#3})!{#5}!({#4})$);
            \node[darkmsgcircle, xshift=5.0mm] at (anchor) {#6};
            \node[xshift=1.5mm] at (anchor) {$\uparrow$};
      }{}
}

\newcommand{\bwmsg}[6]{
      \ifthenelse{\isin{#1}{left} \AND \isin{#2}{down}}{
            \coordinate (anchor) at ($({#3})!{#5}!({#4})$);
            \node[msgdoublecircle, xshift=-5.5mm] at (anchor) {#6};
            \node[xshift=-1.5mm] at (anchor) {$\downarrow$};
      }{}
      \ifthenelse{\isin{#1}{right} \AND \isin{#2}{down}}{
            \coordinate (anchor) at ($({#3})!{#5}!({#4})$);
            \node[msgdoublecircle, xshift=5.5mm] at (anchor) {#6};
            \node[xshift=1.5mm] at (anchor) {$\downarrow$};
      }{}

      \ifthenelse{\isin{#1}{down} \AND \isin{#2}{right}}{
            \coordinate (anchor) at ($({#3})!{#5}!({#4})$);
            \node[msgdoublecircle, yshift=-6.0mm] at (anchor) {#6};
            \node[yshift=-2.0mm] at (anchor) {$\rightarrow$};
      }{}
      \ifthenelse{\isin{#1}{up} \AND \isin{#2}{right}}{
            \coordinate (anchor) at ($({#3})!{#5}!({#4})$);
            \node[msgdoublecircle, yshift=6.0mm] at (anchor) {#6};
            \node[yshift=2.0mm] at (anchor) {$\rightarrow$};
      }{}

      \ifthenelse{\isin{#1}{down} \AND \isin{#2}{left}}{
            \coordinate (anchor) at ($({#3})!{#5}!({#4})$);
            \node[msgdoublecircle, yshift=-6.0mm] at (anchor) {#6};
            \node[yshift=-2.0mm] at (anchor) {$\leftarrow$};
      }{}
      \ifthenelse{\isin{#1}{up} \AND \isin{#2}{left}}{
            \coordinate (anchor) at ($({#3})!{#5}!({#4})$);
            \node[msgdoublecircle, yshift=6.0mm] at (anchor) {#6};
            \node[yshift=2.0mm] at (anchor) {$\leftarrow$};
      }{}

      \ifthenelse{\isin{#1}{left} \AND \isin{#2}{up}}{
            \coordinate (anchor) at ($({#3})!{#5}!({#4})$);
            \node[msgdoublecircle, xshift=-5.5mm] at (anchor) {#6};
            \node[xshift=-1.5mm] at (anchor) {$\uparrow$};
      }{}
      \ifthenelse{\isin{#1}{right} \AND \isin{#2}{up}}{
            \coordinate (anchor) at ($({#3})!{#5}!({#4})$);
            \node[msgdoublecircle, xshift=5.5mm] at (anchor) {#6};
            \node[xshift=1.5mm] at (anchor) {$\uparrow$};
      }{}
}

\newcommand{\bwdarkmsg}[6]{
      \ifthenelse{\isin{#1}{left} \AND \isin{#2}{down}}{
            \coordinate (anchor) at ($({#3})!{#5}!({#4})$);
            \node[darkmsgdoublecircle, xshift=-5.5mm] at (anchor) {#6};
            \node[xshift=-1.5mm] at (anchor) {$\downarrow$};
      }{}
      \ifthenelse{\isin{#1}{right} \AND \isin{#2}{down}}{
            \coordinate (anchor) at ($({#3})!{#5}!({#4})$);
            \node[darkmsgdoublecircle, xshift=5.5mm] at (anchor) {#6};
            \node[xshift=1.5mm] at (anchor) {$\downarrow$};
      }{}

      \ifthenelse{\isin{#1}{down} \AND \isin{#2}{right}}{
            \coordinate (anchor) at ($({#3})!{#5}!({#4})$);
            \node[darkmsgdoublecircle, yshift=-6.0mm] at (anchor) {#6};
            \node[yshift=-2.0mm] at (anchor) {$\rightarrow$};
      }{}
      \ifthenelse{\isin{#1}{up} \AND \isin{#2}{right}}{
            \coordinate (anchor) at ($({#3})!{#5}!({#4})$);
            \node[darkmsgdoublecircle, yshift=6.0mm] at (anchor) {#6};
            \node[yshift=2.0mm] at (anchor) {$\rightarrow$};
      }{}

      \ifthenelse{\isin{#1}{down} \AND \isin{#2}{left}}{
            \coordinate (anchor) at ($({#3})!{#5}!({#4})$);
            \node[darkmsgdoublecircle, yshift=-6.0mm] at (anchor) {#6};
            \node[yshift=-2.0mm] at (anchor) {$\leftarrow$};
      }{}
      \ifthenelse{\isin{#1}{up} \AND \isin{#2}{left}}{
            \coordinate (anchor) at ($({#3})!{#5}!({#4})$);
            \node[darkmsgdoublecircle, yshift=6.0mm] at (anchor) {#6};
            \node[yshift=2.0mm] at (anchor) {$\leftarrow$};
      }{}

      \ifthenelse{\isin{#1}{left} \AND \isin{#2}{up}}{
            \coordinate (anchor) at ($({#3})!{#5}!({#4})$);
            \node[darkmsgdoublecircle, xshift=-5.5mm] at (anchor) {#6};
            \node[xshift=-1.5mm] at (anchor) {$\uparrow$};
      }{}
      \ifthenelse{\isin{#1}{right} \AND \isin{#2}{up}}{
            \coordinate (anchor) at ($({#3})!{#5}!({#4})$);
            \node[darkmsgdoublecircle, xshift=5.5mm] at (anchor) {#6};
            \node[xshift=1.5mm] at (anchor) {$\uparrow$};
      }{}
}

\makeatletter
\DeclareRobustCommand{\cev}[1]{%
  {\mathpalette\do@cev{#1}}%
}
\newcommand{\do@cev}[2]{%
  \vbox{\offinterlineskip
    \sbox\z@{$\m@th#1 x$}%
    \ialign{##\cr
      \hidewidth\reflectbox{$\m@th#1\vec{}\mkern4mu$}\hidewidth\cr
      \noalign{\kern-\ht\z@}
      $\m@th#1#2$\cr
    }%
  }%
}

\def\BibTeX{{\rm B\kern-.05em{\sc i\kern-.025em b}\kern-.08em
    T\kern-.1667em\lower.7ex\hbox{E}\kern-.125emX}}
    
\begin{document}

\title{Variational Bayes for Robust Radar\\ Single Object Tracking}

\author{\IEEEauthorblockN{Alp Sarı}
\IEEEauthorblockA{
\textit{TU Eindhoven}\\
Eindhoven, Netherlands
\\ a.sari@student.tue.nl
}
\and
\IEEEauthorblockN{Tak Kaneko}
\IEEEauthorblockA{
\textit{Sioux Technologies}\\
Eindhoven, Netherlands}
\and
\IEEEauthorblockN{Lense H.M. Swaenen}
\IEEEauthorblockA{
\textit{Sioux Technologies}\\
Eindhoven, Netherlands}
\and
\IEEEauthorblockN{Wouter M. Kouw}
\IEEEauthorblockA{
\textit{TU Eindhoven}\\
Eindhoven, Netherlands\\
w.m.kouw@tue.nl}
}

\IEEEpubid{978-1-6654-8524-1/22/\$31.00 \copyright2022 IEEE}

\maketitle


\vspace{-10px}

\begin{abstract}
We address object tracking by radar and the robustness of the current state-of-the-art methods to process outliers.
The standard tracking algorithms extract detections from radar image space to use it in the filtering stage. Filtering is performed by a Kalman filter, which assumes Gaussian distributed noise. However, this assumption does not account for large modeling errors and results in poor tracking performance during abrupt motions. 
We take the Gaussian Sum Filter (single-object variant of the Multi Hypothesis Tracker) as our baseline and propose a modification by modelling process noise with a distribution that has heavier tails than a Gaussian. Variational Bayes provides a fast, computationally cheap inference algorithm.
%
Our simulations show that - in the presence of process outliers - the robust tracker outperforms the Gaussian Sum filter when tracking single objects. 
\end{abstract}


\section{Introduction}
Radar systems detect objects by emitting electromagnetic waves in the radio spectrum and capturing the signal reflected off objects. They are a critical part of Autonomous Driving (AD) systems for their robustness in varying weather and lighting conditions \cite{vargas_overview_2021}.
Traditional radar tracking algorithms have a processing chain: the detection stage extracts signal intensity peaks from raw measurements, the clustering stage replaces groups of peaks with points, the data association stage assigns these points to objects of interest (e.g., vehicles, pedestrians) and the filtering stage refines the state estimates of the objects \cite{manjunath_radar_2018}. 
We propose an alteration to the filtering stage to increase the robustness to modelling error.

Single object tracking algorithms include the Nearest Neighbour filter, the Probabilistic Data Association filter and the Gaussian Sum Filter \cite{bar-shalom_multitarget-multisensor_1995}.
Unlike Nearest Neighbour and Probabilistic Data Association, the Gaussian Sum Filter considers multiple hypotheses, which improves performance at the cost of increased computational complexity. 
All of these use a Kalman filter during the filtering stage%
, which is the optimal estimator 
in the presence of additive Gaussian process and measurement noise \cite{masnadi-shirazi_step_2019}\cite{ho_bayesian_1964}. 
%
Unfortunately, it is not optimal in the presence of non-Gaussian noise. For example, consider an automotive radar setting that assumes that other cars behave according to constant velocity or constant acceleration dynamics. Sudden movements, such as emergency brakes, generate state evolutions that lie far out in the tails of the Gaussian noise distribution. The probabilities of these points are underestimated which affects the subsequent steps in reasoning over multiple hypotheses, ultimately degrading performance.
These outliers are modelling errors, arising from the difference between the system's behavior and the model's assumptions. 
This paper proposes an alternate noise model, namely the heavy-tailed Student's t-distribution, to avoid underestimation of outlying state dynamics. Specifically, we replace the Kalman filter block of Gaussian Sum Filter with the Robust Student's t Kalman filter \cite{huang_novel_2017}, which increases the overall algorithm's robustness to model errors.
 
%
Our key contributions are: a demonstration of the GSF's susceptibility to noise model error, a model specification and inference algorithm using a Student's t-distribution noise model, and a performance comparison between the proposed algorithm and the GSF. Simulated experiments show that the resulting filter is more robust to noise model errors.

\section{Problem Setting}
\IEEEpubidadjcol
Consider an object with state $x_k \in \mathbb{R}^{D_x}$ that generates a measurement $o_k \in \mathbb{R}^{D_o}$, corresponding to a peak in signal power. The detection stage finds the peak - along with peaks produced by noise - and outputs a series of point detections $Z_k = \{z_k^{i} \in \mathbb{R}^{D_o} : i = 1, \dots, m_k \}$. Note that it is also possible that the object was not detected by the peak-finder, in which case all $Z_k$ correspond to noise peaks, a.k.a. \emph{clutter}.
%
%
Our goal is to recursively infer the state distribution given all detections so far, i.e., estimate $p(x_k \mid Z_{1:k})$ \cite{sarkka_bayesian_2013}.


To indicate which detection corresponds to the object measurement, we introduce a data association variable $\theta_k \in \{0,1,\dots,m_k\}$ such that $z^{\theta_k}_k = o_k$. Our likelihood then becomes a joint probability over data associations and detections, i.e., $p(\theta_k, Z_k \given x_k)$. Using the Bayesian filtering equations \cite[Ch.~4]{sarkka_bayesian_2013}, we may write:
\begin{align} \label{eq:bayes}
    p\left(x_k \given \theta_{1:k}, Z_{1:k} \right) = \frac{p\left(\theta_k, Z_k \, | \, x_k \right)p\left( x_k \, | \,  \theta_{1:k\-1}, Z_{1:k\-1}\right)}{p\left(\theta_k, Z_k \given \theta_{1:k\-1}, Z_{1:k\-1} \right)} ,
\end{align}
where the numerator consists of the likelihood times the prior predictive distribution and the denominator is the evidence term. The special case $\theta_k = 0$ indicates that none of the detections correspond to the object. In that case, the observations are not dependent on the object, $p(\theta_k = 0, Z_k \given x_k) = p(\theta_k = 0, Z_k)$, and the state posterior in Equation \eqref{eq:bayes} will be proportional to the prior predictive, a situation equivalent to filtering with "missing" observations.

Note that Equation \eqref{eq:bayes} is an intermediate posterior distribution as it still depends on the data associations. To obtain the exact state posterior, we must marginalize over $\theta_{1:k}$. This marginalization corresponds to weighting each possible data association, i.e., each possible sequence of object evolutions, with their probability: 
\vspace{-10pt}
\begin{align}
    p\big(x_k &\mid Z_{1:k} \big) = \nonumber \\
    &\sum_{\theta_1} ... \sum_{\theta_k} p\left(x_k \mid \theta_{1:k}, Z_{1:k}\right) \Pr\big[\theta_{1:k} \mid Z_{1:k}\big]
\end{align}
where $\Pr[\cdot]$ refers to a probability mass function. But here we find a problem: the number of hypothetical sequences is $\prod_{j=1}^{k}(m_j+1)$. Calculating the exact posterior quickly becomes intractable, which indicates a need for approximation.


\subsection{Gaussian Sum Filter}
The Gaussian Sum Filter (GSF) approximates the posterior distribution by a Gaussian mixture model with fewer components than the original mixture. 
The GSF models the state transition $p(x_k \given x_{k-1})$ and measurement model of a single detection $p(z_k^{\theta_k}\given x_k)$ by a linear state space model with additive Gaussian noise. Clutter detections are modeled with a Poisson Point Process (Poisson Random Finite Set) with an intensity function $\lambda_c(\cdot)$ \cite{vo_bayesian_2008}. Then, the conditional posterior distribution $p\left(x_k \mid \theta_k,Z_k\right)$ is calculated using the Kalman filter (KF) equations. Also, the calculation of the mixture weights $\Pr\left[\theta_k \mid Z_k\right]$ requires the evidence term $p(z_k^{\theta_k})$, which has a tractable expression, given model specifications above. 

After obtaining the conditional posteriors and weights, the GSF performs a mixture reduction step, which is a complex problem in itself. We choose the following simple approach: We cap the maximum number of components at $N_{\text{max}}$ and prune the components with weights smaller than a threshold $\gamma_{\text{prune}}$. If we are left with more than $N_{\text{max}}$ components, Runnall's algorithm \cite{runnalls_kullback-leibler_2007} is used to reduce to mixture to $N_{\text{max}}$ components.

\subsection{Susceptibility to Process Outliers}
\label{sec:ProblemGSF}
\begin{figure}[thb]
    \centering
    \begin{adjustbox}{max size={\textwidth/2}{\textheight/2}}
    \input{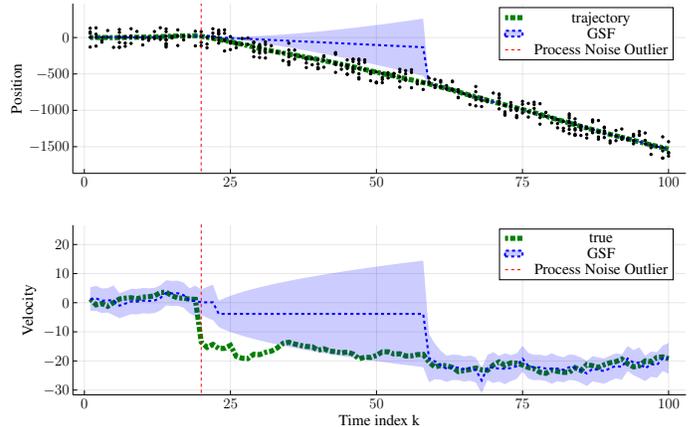}
    \end{adjustbox}
    \vspace*{-3mm}
    \caption{A simulated tracking scenario. Position~(top), velocity~(bottom) and corresponding confidence interval estimates over time of GSF~(blue-dashed) versus true trajectory~(green-dashed). Black dots refer to detections~(including clutter). The vertical line~(red) indicates when the process outlier is introduced.}
    \label{fig:GSFProblemx}
\end{figure}

The GSF has a zero mean Gaussian noise model for the process noise, which accounts for the mismatch between the modeled dynamics of the target and the true motion dynamics. Note that the Gaussian distribution has slim tails, and $99.7\%$ of the probability mass falls into the $\pm 3$ standard deviation around the mean value. Thus, the mismatch between the model and true dynamics is explained well by the Gaussian noise if the error is in $\pm 3$ standard deviation.
But, when the target moves differently than the expected motion dynamics, the error increases and the error values fall into the tails of the Gaussian, which are referred as outliers. When an outlier is encountered, the Gaussian distributed process noise underestimates its probability density. An illustration of the problem is shown in Fig.~\ref{fig:GSFProblemx}. We simulate the observation of positions in clutter. The observed detections are shown with black dots in the position plots (top). The blue dashed line is the state estimate using GSF and the corresponding ribbon indicates the $\pm 3$ standard deviation of the state estimate. At the $20^{th}$ time step (indicated by a vertical red dashed line), the state transition is corrupted with an outlier value and it is observed that the state estimation drifts off.

\wouter{I think the connection between Sections II and III are not clear enough. We talk about $Z_k$ in Sec. II and about $o_k$ in Sec. III.}

\section{Robust Extension of GSF}
We will model both the measurement and the predictive prior distribution with a Student's t-distribution to address the susceptibility to process outliers (Sec.~\ref{sec:ProblemGSF}). The Student's t-distribution has heavier tails than the Gaussian distribution, but inference using the Student's t is intractable.
\wouter{Possible addition: reference to Gaussian scale mixture to motivate use of Student's t more elaborately.}
We perform approximate inference using variational Bayes (VB). The resulting filtering algorithm is known as the Robust Student's t Kalman filter (RSTKF). The RSTKF and VB update equations have previously been derived in \cite{huang_novel_2017}, and this paper is an extension towards single object tracking using the GSF algorithm.

\subsection{Model Specification}
\label{sec:StateSpace}
Consider a discrete-time linear state-space model with additive noise expressed as:
 \begin{subequations}\label{eq:LinearSSM2}
   \begin{align}
        x_k &= F x_{k\-1} + \omega_k \label{eq:StateTransition} \\
        o_k &= H x_{k} + \nu_k \label{eq:Likelihood}\, ,
   \end{align}
\end{subequations}
where $F$ is the state transition matrix, $H$ is the observation matrix, and $\omega_k$ and $\nu_k$ are referred as process and measurement noise, respectively. In probabilistic form, this becomes:
\begin{subequations} 
\label{eq:probssm}
\begin{align}
    p(x_k \mid x_{k-1}) &= p_{\omega_k}(x_k - F x_{k-1}) \\
    p(o_k \mid x_{k}) &= p_{\nu_k}(o_k - H x_{k}) \, ,
\end{align}
\end{subequations}
where $p_{\omega_k}(\cdot)$ and $p_{\nu_k}(\cdot)$ are the process noise and measurement noise pdfs, respectively.
Using \eqref{eq:StateTransition}, the predictive prior is obtained through the Chapman-Kolmogorov equation which in this case yields \cite[Ch.~3]{sarkka_bayesian_2013}:
\begin{align} \label{eq:PredictivePrior}
    p\big(x_k &\given o_{1:k\-1} \big) \nonumber \\
    &= \! \int p_{\omega_k}\left(x_k - F x_{k\-1} \right) p\left(x_{k\-1} \given o_{1:k\-1} \right)\mathrm{d}x_{k\-1}\, .
\end{align}


A standard KF assumes Gaussian noise distributions, with zero mean vectors and covariance matrices $Q$ and $R$, respectively. Then, the predictive prior and likelihood terms are computed as:
 \begin{subequations}
 \label{eq:KFModel}
  \begin{align}
    p\big(x_k \given o_{1:k\-1} \big) &= \mathcal{N}\left(x_k\given \mu_{k\given k\-1 },P_{k\given k\-1}\right) \\
    p\big(o_k \given x_{k} \big) &= \mathcal{N}\left(o_k\given Hx_k,R\right)\, ,
 \end{align}
 \end{subequations}
where $\mu_{k\given k\-1}$ and $P_{k\given k\-1}$ are:
\begin{align}
    \mu_{k\given k\-1} = F \mu_{k\-1\given k\-1} \, , \quad
    P_{k\given k\-1} = F P_{k\-1 \given k\-1} F^T + Q \, .    \label{eq:MeanCov}
\end{align}
Instead of a Gaussian, we adopt a Student's t-distribution for the predictive prior and likelihood, which gives:
\begin{subequations}
    \label{eq:RSTKFModelSpec}
\begin{align}
    p\big(x_k \given o_{1:k\-1},\Sigma_k \big)&= \mathcal{T}\big(x_k\given \mu_{k\given k\-1 },\Sigma_k,s\big) \\ 
    p\big(o_k \given x_{k},\big) &= \mathcal{T}\big(o_k\given H x_{k},R,v \big)\, ,
\end{align}
\end{subequations}
where $\mathcal{T}(x \given \mu,\Lambda,s)$ denotes a generalized Student's t-distribution with location $\mu$, scale matrix $\Lambda$ and degrees of freedom $s$. We treat
scale matrix of the prior term $\Sigma_k$ as an unknown random variable, which increases the uncertainty factored into the model.
\lense{Is the following view correct?: ``We could also create a model like $x_k = Fx_{k-1} + \omega_k$ with $\omega_k \sim \mathcal{T}\big(0, \Sigma_k, s\big)$, but we wouldn't get the same clean analytical expressions and inference procedure out of it.'' Should we add any such comment?}
\alp{Yes, it is correct. If you specify $\omega_k$ parametric but other than a Gaussian, \eqref{eq:PredictivePrior} won't have an analytical solution.}

Aside from being a distribution with a heavier tail, the Student's t-distribution can also be expressed as an infinite mixture of Gaussians with the same mean but different scales:
\begin{subequations}
\begin{alignat}{2}    
    &\mathcal{T} \big( x_k &&\given \mu_{k\given k\-1 },\Sigma_k,s \big) \nonumber\\
    & &&= \int \mathcal{N}\left(x_k \given \mu_{k\given k\- 1}, \frac{\Sigma_k}{\xi_k} \right) \mathcal{G}\left(\xi_k \given\frac{s}{2}, \frac{s}{2} \right)\mathrm{d} \xi_{k} \label{eq:InfiniteGaussPrior} \\
    &\mathcal{T} \big( o_k &&\given H x_{k},R,v \big) \nonumber\\
    & &&= \int \mathcal{N}\left(o_k \given H x_{k}, \frac{R}{\lambda_k} \right) \mathcal{G}\left(\lambda_k \given\frac{v}{2}, \frac{v}{2} \right)\mathrm{d} \lambda_{k} \label{eq:InfiniteGaussMeas} \, ,
\end{alignat}
\end{subequations}
where $\mathcal{G}(.\given\alpha,\beta)$ denotes a Gamma distribution with shape parameter $\alpha$ and rate parameter $\beta$, $\xi_k$ and $\lambda_k$ are scale parameters. Infinite mixture representation shows that model  specification in \eqref{eq:RSTKFModelSpec} is richer than \eqref{eq:KFModel}, which only has a single Gaussian expression \alp{better wording is possible}. 
Moreover, the model specification can be written as a hierarchical state space model using  \eqref{eq:InfiniteGaussPrior} and \eqref{eq:InfiniteGaussMeas}. 
The predictive prior of the hierarchical state space model is\alp{Noting that we write $p\left(x_k \given \Sigma_k, \xi_k,o_{1:k\-1}\right)$ here whereas the last time the prior was mentioned, it was without $\xi_k$,i.e., $p\left(x_k \given \Sigma_k,o_{1:k\-1}\right)$. Is this a leap that we need to explain?}:
\begin{subequations}
\label{eq:RSTKFStateGauss2}
\begin{align}
    &p\left(x_k \given \Sigma_k, \xi_k,o_{1:k\-1}\right) =  \mathcal{N}\left(x_k \given \mu_{k},\Sigma_k/\xi_k\right)\\
    &p\left(\Sigma_k\right) =  \mathcal{IW}\left(\Sigma_k \given u_k,U_k\right)\\
    &p\left(\xi_k\right) =  \mathcal{G}\left(\xi_k \given \frac{s}{2},\frac{s}{2}\right)\, ,
\end{align}
\end{subequations}
where $\mathcal{IW}\left(\Sigma_k \given u_k,U_k\right)$ denotes the inverse-Wishart distribution with $u_k$ degrees of freedom and scale matrix $U_k$. The scale matrix is set as $U_k= P_{k|k\-1}\cdot(u_k - D_x -1)$ to ensure $\mathbb{E}[\Sigma_k] = P_{k|k\-1}$. The likelihood hierarchical model is:
\begin{subequations}
\label{eq:RSTKFMeasGauss2}
\begin{align}
    &p\left(o_k \given \lambda_k,x_k\right) =  \mathcal{N}\left(o_k \given H x_k,R_k/\lambda_k\right)\\
    &p\left(\lambda_k\right) =  \mathcal{G}\left(\lambda_k \given \frac{v}{2},\frac{v}{2}\right) \, .
\end{align}
\end{subequations}
Note that the model specification in \eqref{eq:RSTKFStateGauss2}  and \eqref{eq:RSTKFMeasGauss2} is a special case of Gaussian Scale Mixture (GSM) distribution. More flexibility, such as skewed heavy-tailed distributions, can be introduced using this framework. For more details about GSM distributions, we refer the interested reader to \cite{huang_robust_2019}.

Unfortunately, with the given model specification, computing the posterior $p(x_k,\Sigma_k,\xi_k,\lambda_k|o_{1:k})$ exactly is intractable.

\subsection{Inference using Variational Bayes}
Variational Bayes refers to adopting a second probabilistic model, called the \emph{variational} model $q$, and approximating the original model using the calculus of variations \cite{blei_variational_2017}. The variational model is optimized through the \emph{free energy} functional at time step $k$:
\begin{align} \label{eq:FreeEnergy1}
     \F_k[q] \! \triangleq \mathbb{E}_q \big[\ln \! \frac{q(x_k,\Sigma_k,\xi_k,\lambda_k)}{p(x_k,\Sigma_k,\xi_k,\lambda_k | o_{1:k})} \big]\! - \! \ln p(o_k | o_{1:k\-1}) .
\end{align}
The first term is a Kullback-Leibler (KL) divergence between the variational model and the posterior distribution, indicating the quality of the approximation. However, evaluating this expression requires the true posterior distribution. This problem can be circumvented by applying Bayes' rule:
\begin{align} \label{eq:FreeEnergy2}
     \F_k[q] = \mathbb{E}_q \big[\ln &\frac{q(x_k,\Sigma_k,\xi_k, \lambda_k)}{p(x_k,\Sigma_k,\xi_k | o_{1:k\-1})} \big] \nonumber \\
     & \qquad \qquad \qquad - \mathbb{E}_q \big[\ln p(o_k, \lambda_k | x_k) \big] \, ,
\end{align}
where the posterior times the evidence was decomposed into the prior, corresponding to \eqref{eq:RSTKFStateGauss2}, and the likelihood, corresponding to \eqref{eq:RSTKFMeasGauss2}.

The minimisation of the free energy functional requires the calculus of variations, but we can obtain a simplified problem by parameterizing the variational model $q$ \cite{sarkka2009recursive} \wouter{I've used a paper by Särkkä on simultaneous state and noise estimation using Variational Bayes. A good alternative might be just a textbook that contains an explanation of variational Bayes, such Bishop's book (chapter 10) or Murphy's (new) book}. First, we assume it factorizes according to:
\begin{align}
    q(x_k,\Sigma_k,\xi_k,\lambda_k) = q_x(x_k)\, q_{\Sigma}(\Sigma_k) \, q_{\xi}(\xi_k)\, q_{\lambda}(\lambda_k) \, .
\end{align}
Then, we impose the following parametric distributions:
\begin{subequations}
\begin{align}
    q_x(x_k) &\triangleq \mathcal{N}(x_k \mid m_k, S_k) \\
    q_{\Sigma}(\Sigma_k) &\triangleq \mathcal{IW}(\Sigma_k \mid \Lambda_k, \nu_k) \\
    q_{\xi}(\xi_k) &\triangleq \mathcal{G}(\xi_k \mid \alpha_k, \beta_k) \\
    q_{\lambda}(\lambda_k) &\triangleq \mathcal{G}(\lambda_k \mid \gamma_k, \delta_k) \, .
\end{align}
\end{subequations}
The free energy functional is now an objective function with respect to $\phi =\{m_k, S_k, \Lambda_k, \nu_k, \alpha_k, \beta_k, \gamma_k, \delta_k \}$. The optimal form of each factor consists of a marginalization with respect to the other factors \cite{blei_variational_2017}. The optimal form of the state factor is:
\begin{align}
    q_x(x_k) &\propto \exp \big(\mathbb{E}_{q_\Sigma, q_\xi} \big[ \log p(x_k, \Sigma_k, \xi_k \given o_{1:k\-1}) \big] \nonumber \\
    &\qquad \qquad \qquad \qquad + \mathbb{E}_{q_\lambda} \big[ \log p(o_k, \lambda_k \given x_k) \big] \big) \, ,
\end{align}
which - when solved - yields the following update equations for its parameters:
\begin{subequations}
\begin{align}
    m_k &= \mu_{k\given k\-1} + K_k \big(o_k - H_k \mu_{k\given k\-1} \big) \\
    S_k &= \tilde{S}_k - K_k H_k \tilde{S}_k \, ,
\end{align}
\end{subequations}
with Kalman gain $K_k = \tilde{S}_k H_k^{\top}(H_k \tilde{S}_k H_k^{\top} + \tilde{R}_k)$. The auxiliary matrices $\tilde{S}_k$ and $\tilde{R}_k$ correspond to the expectations of the scaled covariance matrix of the prior predictive and the scaled covariance of the likelihood:
\begin{align}
    \tilde{S}_k = \frac{1}{\nu_k - D_x -1}\Lambda_k \frac{\beta_k}{\alpha_k} \, , \qquad 
    \tilde{R}_k = R_k \frac{\delta_k}{\gamma_k} \, .
\end{align}
For a more detailed description of the derivations involved in, see \cite{huang_novel_2017}. Note that the covariance matrix $S_k$ of the factor is now time-varying, as it depends on the current belief over the process noise covariance matrix $\Sigma_k$. 
\alp{I changed the ordering of the sentence from "Note that the covariance matrix of the recognition factor $S_k$ ..." to "Note that the covariance matrix $S_k$ of the factor ... as I think the factor at hand is the state factor. Is this true?"}
Solving for the other factors is more straightforward as they only appear in term each. The inverse-Wishart distributed factor for $\Sigma_k$ has the following optimal form:
\begin{align}
    q_\Sigma(\Sigma_k) &\propto \exp \big(\mathbb{E}_{q_x, q_\xi} \big[ \log p(x_k, \Sigma_k, \xi_k \given o_{1:k\-1}) \big] \big) \, ,
\end{align}
which gives the parameter update equations:
\begin{align}
    \Lambda_k = U_k + \frac{\alpha_k}{\beta_k} \mathbb{C}_x \, , \quad
    \nu_k = u_k + 1
\end{align}
where $\mathbb{C}_x = S_k + (m_k - \mu_{k\given k\-1})(m_k - \mu_{k\given k\-1})^{\top}$.
The factor for the process scale parameter $\xi_k$ has an optimal form of:
\begin{align}
    q_\xi(\xi_k) &\propto \exp \big(\mathbb{E}_{q_\Sigma, q_x} \big[ \log p(x_k, \Sigma_k, \xi_k \given o_{1:k\-1}) \big] \big) \, ,
\end{align}
yielding parameter updates:
\begin{align}
    \alpha_k \! = \! \frac{1}{2}(D_x \! + \! s) , \
    \beta_k \! = \! \frac{1}{2}\big(s \! + \! \text{tr}(\mathbb{C}_x (\nu_k \! + \! D_x \! - \! 1)\Lambda_k^{-1} \big) \, .
\end{align}
Finally, the factor for the measurement noise scale parameter $\lambda_k$ is optimal when:
\begin{align}
    q_\lambda(\lambda_k) &\propto \exp \big(\mathbb{E}_{q_x} \big[ \log p(o_k, \lambda_k \given x_k) \big] \big) \, ,
\end{align}
yielding parameter update equations:
\begin{align}
    \gamma_k = \frac{1}{2}(D_{o} + v ) \, , \quad
    \delta_k = \frac{1}{2}\big(v + \text{tr}(\mathbb{C}_o R_k^{-1}) \big) \, ,
\end{align}
where $\mathbb{C}_o = (o_k - H m_k)(o_k - H m_k)^T + H S_k H^{\top}$.
These update equations are a form of exact coordinate descent on $\mathcal{F}_k[\phi]$ \cite{blei_variational_2017}.
It typically takes less than 10 iterations to reach convergence.


\subsection{Robust Gaussian Sum Filter Implementation}
%
The desired robust sum filter is achieved simply by replacing the KF block with RSTKF block, which is responsible of calculating the conditional posterior distribution $p(x_k | Z_k,\theta_k)$. The resulting algorithm will be referred as \textit{RSTKF-GSF}.
Swapping the KF block affects the computation of the mixture weights since these require the calculation of $p(z_k^{\theta_k})$. However, that term becomes intractable. We approximate it using the formula for weight computation under the standard GSF. 
%

\begin{figure}[h!]
    \centering
    \begin{adjustbox}{max size={\textwidth/2}{\textheight/2}}
    \input{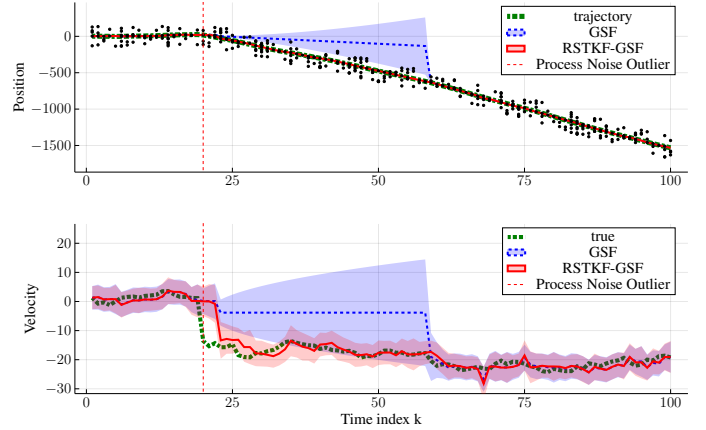}
    \end{adjustbox}
    \vspace*{-3mm}
    \caption{Same simulation as in Fig.~\ref{fig:GSFProblemx}, including the estimates using \textit{RSTKF-GSF}(red-solid). The performance of \textit{RSTKF-GSF} does not degrade substantially after encountering the outlier process noise.}
    \label{fig:GSFProblemRSTKFx}
\end{figure}

Fig.~\ref{fig:GSFProblemRSTKFx} shows the performance of the \textit{RSTKF-GSF} in the same scenario as Fig.~\ref{fig:GSFProblemx}. It is robust to the sudden large drop in velocity that caused the GSF to start drifting.

\vspace{-10pt}
\section{Experiments}
\label{sec:Experiments}
In this section, we compare the performance of RSTKF-GSF with the original GSF algorithm on a simulated scenario for single object tracking in clutter. Four different Monte Carlo simulations were performed, where the process and the measurement noise are both Gaussian in the first scenario. The second scenario simulates heavy-tailed process noise and Gaussian measurement noise. The third scenario simulates Gaussian process noise and heavy-tailed measurement noise. The fourth scenario simulates both heavy-tailed process and measurement noise.

\subsection{Simulation Setting}
The simulation setting and performance metric is similar to given in \cite{huang_novel_2017}. Each Monte Carlo simulation consist of $100$ time steps. The trajectory is simulated with a single target moving according to a constant velocity model in 2-D space and the position of the target is observed. The state at time step $k$ is defined as $x_k \triangleq \left[p_{x,k}, \, p_{y,k}, \, v_{x,k}, \, v_{y,k} \right]^\top$ , where $(p_{x,k}, \, p_{y,k})$ is the position of the target and $(v_{x,k}, \, v_{y,k})$ is the velocity of the target in Cartesian coordinates in x-axis and y-axis, respectively. The linear state space model is used given in \eqref{eq:LinearSSM2}. The parameters of the dynamics are set as follows:
\begin{align}
    F = \begin{bmatrix}
    I_2       & \Delta_t I_2   \\
    \mathbf{0}_2    & I_2 \\
\end{bmatrix} \, , \quad
Q = \begin{bmatrix}
    \frac{\Delta_t^3}{3}I_2 & \frac{\Delta_t^2}{2}I_2 \\
    \frac{\Delta_t^2}{2}I_2 & \Delta_t I_2 \\
\end{bmatrix} , 
\end{align}
where $I_2$ is the 2-D identity matrix, $\mathbf{0}_2$ is the $2x2$ matrix of zeros, $\Delta_t =1$ is the sampling rate.
The parameters of the measurement process are:
\begin{align}
H = \begin{bmatrix}
    I_2& \mathbf{0}_2 \\
\end{bmatrix} \, , \quad   R = r I_2 \, ,
\end{align}

where $r=10$. The process model is simulated according to:
\begin{align}
    \omega_k &\sim
        \left\{ \begin{array}{ll}
            \mathcal{N}(0,Q) &\text{w.p. $P_\omega$} \\
            \mathcal{N}(0,100\, Q) &\text{w.p. $1- P_\omega$}
        \end{array} \right . \, ,
\end{align}
where "w.p." is shorthand for "with probability". The measurement model is simulated with:
\begin{align}
    v_k &\sim
    \left\{ \begin{array}{ll}
        \mathcal{N}(0,R) &\text{w.p. $P_v$} \\
        \mathcal{N}(0,100 \, R) &\text{w.p. $1- P_v$}
    \end{array} \right. \, .
\end{align}
\wouter{todo: add sentence that choices of $P_\omega$ and $P_v$ indicate heavy-tailed noise.}
The set of parameters for each experiment case is summarized in Table.~\ref{table:ExperimentSetting}. Chosen values of $P_\omega$ and $P_v$ are used to simulate heavy-tailed noise~\cite{huang_novel_2017}.
\begin{table}[h!]
\setlength{\tabcolsep}{8pt}
\caption{$P_\omega$ and $P_v$ parameters of experiments.}
\begin{tabular}{r | l | l | r | r}
                                      Exp. \# & Process Noise & Measurement Noise & $P_\omega$ & $P_v$ \\
                                       \midrule
1 & Gaussian & Gaussian      & 1.00       & 1.00           \\
2 & Heavy-Tailed & Gaussian    & 0.95    & 1.00           \\
3 & Gaussian & Heavy-Tailed    & 1.00       & 0.90        \\
4 & Heavy-Tailed & Heavy-Tailed & 0.95    & 0.90      
\end{tabular}
\label{table:ExperimentSetting}
\end{table}

Each experiment consist of $M=1000$ Monte Carlo simulations comparing the Root Mean Square Errors (RMSE) at each time step, for both position and velocity.

The clutter is sampled from a Poisson Point Process with rate parameter $3$. The spatial pdf is uniform around the true state position with a range of $\pm 15r$, i.e., $150$. The probability of detection $P_D$ was set to $0.95$, which is the value used for both simulating the data and for both GSF filters. The clutter intensity function $\lambda_c(\cdot)$ is assumed to be constant and set to $3 / (2\cdot 15 \cdot r)^2$.
Mixture reduction parameters are set as $\gamma_{prune} = 6.25\cdot 10^{-6}$ and $N_{max} = 10$ for both filters. Weight calculations are also identical in order to compare filtering algorithms.


\subsection{Results}
\label{sec:Results}

\begin{figure}[h!]
    \centering
    \begin{adjustbox}{max size={\textwidth/2}{\textheight/2}}
    \input{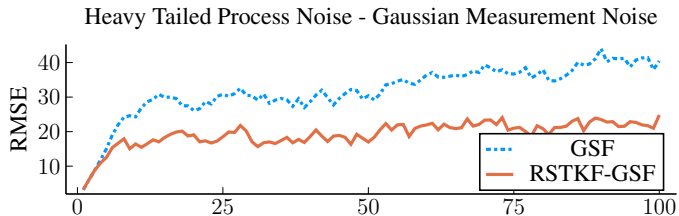}
    \end{adjustbox}
    \vspace*{-3mm}
    \caption{Experiment-2 with heavy-tailed process noise and Gaussian distributed measurement noise. Root mean square errors over time of GSF (blue-dashed) versus RSTKF-GSF (red-solid).}
    \label{fig:Exp2}
\end{figure}

In radar object tracking, cars will not always behave according to constant-velocity model and sudden brakes or accelerations cause outliers in the state transitions. This corresponds to the setting with heavy-tailed process noise and just Gaussian measurement noise. The simulation results of this scenario are shown in Fig.~\ref{fig:Exp2}. They show that the \textit{RSTKF-GSF} outperforms the GSF. This is expected since \textit{RSTKF-GSF} models heavy-tailed noise characteristics whereas GSF assumes Gaussian process noise.

\begin{figure}[h!]
    \centering
    \begin{adjustbox}{max size={\textwidth/2}{\textheight/2}}
    \input{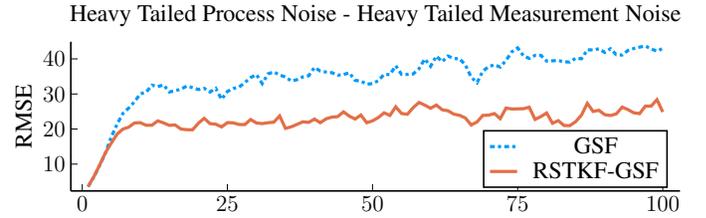}
    \end{adjustbox}
    \vspace*{-3mm}
    \caption{Experiment-4 with heavy-tailed process noise and measurement noise. Root mean square errors over time of GSF (blue-dashed) versus RSTKF-GSF (red-solid).}
    \label{fig:Exp4}
\end{figure}

Similarly, the \textit{RSTKF-GSF} preserves its robustness compared to the GSF when both process and measurement noise are heavy-tailed. This result is shown in Fig.~\ref{fig:Exp4}.
%

\begin{figure}[h!]
    \centering
    \begin{adjustbox}{max size={\textwidth/2}{\textheight/2}}
    \input{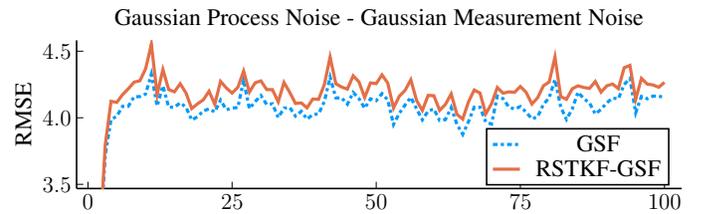}
    \end{adjustbox}
    \vspace*{-3mm}
    \caption{Experiment-1 with Gaussian distributed process noise and measurement noise. Root mean square errors over time of GSF (blue-dashed) versus RSTKF-GSF (red-solid).}
    \label{fig:Exp1}
\end{figure}

\begin{figure}[h!]
    \centering
    \begin{adjustbox}{max size={\textwidth/2}{\textheight/2}}
    \input{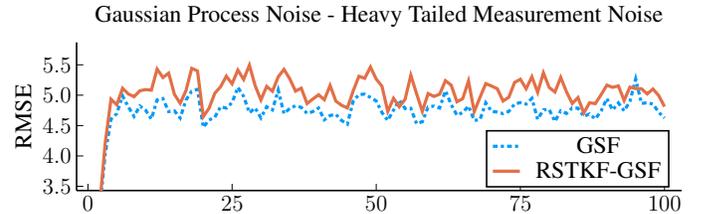}
    \end{adjustbox}
    \vspace*{-3mm}
    \caption{Experiment-3 with Gaussian distributed process noise and heavy-tailed measurement noise. Root mean square errors over time of GSF (blue-dashed) versus RSTKF-GSF (red-solid).}
    \label{fig:Exp3}
\end{figure}

Fig.~\ref{fig:Exp1} and Fig.~\ref{fig:Exp3} show that when the process noise is Gaussian, i.e., no outlier in the process, both filters perform similarly in terms of RMSE.


\subsection{Computational cost}
\begin{figure}[htbp]
    \centering
    \begin{adjustbox}{max size={0.48\textwidth}{\textheight/2}}
    \input{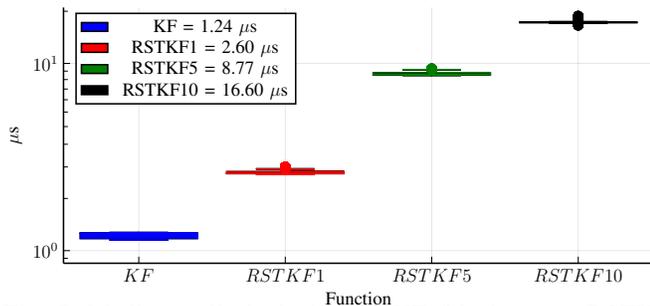}
    \end{adjustbox}
    \vspace*{-3mm}
    \caption{Median wall clock time of KF~(blue) versus RSTKF with $1$~(red), $5$~(green) and $10$~(black) iterations. Results for the inner $90^{th}$ percentile are shown on logarithmic scale.}
    \label{fig:TimeBenchmark}
\end{figure}

The filtering parts of both algorithms are also compared in terms of computation time. The RSTKF performs several parameter updates in each time step to reach convergence. We obtain results in Sec.~\ref{sec:Results} using $10$ iterations per time step, but we observed that $5$ iterations per time step were sufficient for posterior parameters to converge. Fig.~\ref{fig:TimeBenchmark} compares the median wall clock time of RSTKF function with varying number of iterations and the KF function. 
We observe that using the RSTKF algorithm with $10$ iterations does not introduce considerable computational costs, and RSTKF is feasible to use in real-time applications.

\subsection{Adaptive estimation versus tuning}
\alp{denoted scale parameter with $c$, and also changed the x-label of the Figure. Should I also include $c$ in the caption?}
\wouter{Yes. $c$ is necessary to read the figure.}
\alp{Maybe we should mention that this tuning performed for the same data in Fig.1-2?}
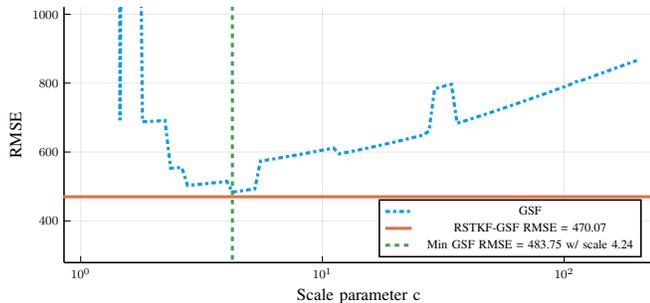
\begin{figure}[htbp]
    \centering
    \begin{adjustbox}{max size={.48\textwidth}{\textheight/2}}
    \begin{tikzpicture}[]
\begin{axis}[point meta max={nan}, point meta min={nan}, legend cell align={left}, legend columns={1}, title={}, title style={at={{(0.5,1)}}, anchor={south}, font={{\fontsize{14 pt}{18.2 pt}\selectfont}}, color={rgb,1:red,0.0;green,0.0;blue,0.0}, draw opacity={1.0}, rotate={0.0}}, legend style={color={rgb,1:red,0.0;green,0.0;blue,0.0}, draw opacity={1.0}, line width={1}, solid, fill={rgb,1:red,1.0;green,1.0;blue,1.0}, fill opacity={1.0}, text opacity={1.0}, font={{\fontsize{8 pt}{10.4 pt}\selectfont}}, text={rgb,1:red,0.0;green,0.0;blue,0.0}, cells={anchor={center}}, at={(0.98, 0.02)}, anchor={south east}}, axis background/.style={fill={rgb,1:red,1.0;green,1.0;blue,1.0}, opacity={1.0}}, anchor={north west}, xshift={1.0mm}, yshift={-1.0mm}, clip mode={individual}, width={150.4mm}, height={74.2mm}, scaled x ticks={false}, xlabel={Scale parameter c}, x tick style={color={rgb,1:red,0.0;green,0.0;blue,0.0}, opacity={1.0}}, x tick label style={color={rgb,1:red,0.0;green,0.0;blue,0.0}, opacity={1.0}, rotate={0}}, xlabel style={at={(ticklabel cs:0.5)}, anchor=near ticklabel, at={{(ticklabel cs:0.5)}}, anchor={near ticklabel}, font={{\fontsize{11 pt}{14.3 pt}\selectfont}}, color={rgb,1:red,0.0;green,0.0;blue,0.0}, draw opacity={1.0}, rotate={0.0}}, xmode={log}, log basis x={10}, xmajorgrids={true}, xmin={0.8530394184621627}, xmax={234.45575394458902}, xtick={{1.0,10.0,100.0}}, xticklabels={{$10^{0}$,$10^{1}$,$10^{2}$}}, xtick align={inside}, xticklabel style={font={{\fontsize{8 pt}{10.4 pt}\selectfont}}, color={rgb,1:red,0.0;green,0.0;blue,0.0}, draw opacity={1.0}, rotate={0.0}}, x grid style={color={rgb,1:red,0.0;green,0.0;blue,0.0}, draw opacity={0.1}, line width={0.5}, solid}, axis x line*={left}, x axis line style={color={rgb,1:red,0.0;green,0.0;blue,0.0}, draw opacity={1.0}, line width={1}, solid}, scaled y ticks={false}, ylabel={RMSE}, y tick style={color={rgb,1:red,0.0;green,0.0;blue,0.0}, opacity={1.0}}, y tick label style={color={rgb,1:red,0.0;green,0.0;blue,0.0}, opacity={1.0}, rotate={0}}, ylabel style={at={(ticklabel cs:0.5)}, anchor=near ticklabel, at={{(ticklabel cs:0.5)}}, anchor={near ticklabel}, font={{\fontsize{11 pt}{14.3 pt}\selectfont}}, color={rgb,1:red,0.0;green,0.0;blue,0.0}, draw opacity={1.0}, rotate={0.0}}, ymajorgrids={true}, ymin={279.0}, ymax={1021.0}, ytick={{400.0,600.0,800.0,1000.0}}, yticklabels={{$400$,$600$,$800$,$1000$}}, ytick align={inside}, yticklabel style={font={{\fontsize{8 pt}{10.4 pt}\selectfont}}, color={rgb,1:red,0.0;green,0.0;blue,0.0}, draw opacity={1.0}, rotate={0.0}}, y grid style={color={rgb,1:red,0.0;green,0.0;blue,0.0}, draw opacity={0.1}, line width={0.5}, solid}, axis y line*={left}, y axis line style={color={rgb,1:red,0.0;green,0.0;blue,0.0}, draw opacity={1.0}, line width={1}, solid}, colorbar={false}]
    \addplot[color={rgb,1:red,0.0;green,0.6056;blue,0.9787}, name path={eddb85ef-cc49-40c5-877b-ade10f33fb21}, draw opacity={1.0}, line width={2}, dashdotted]
        table[row sep={\\}]
        {
            \\
            1.0  19316.696682578055  \\
            1.0549763580417753  19308.5054707385  \\
            1.112975116027088  19297.414229662816  \\
            1.1741624344973793  9087.393248178763  \\
            1.2387136088955097  8451.595946910898  \\
            1.3068135717693687  6123.414127737441  \\
            1.3786574225848127  6121.259156469127  \\
            1.4544509866657864  692.1415842255557  \\
            1.5344114048629378  3325.7939281803265  \\
            1.6187677556400661  4626.889470896407  \\
            1.7077617113606152  2239.6609836322364  \\
            1.801648230654411  687.8209426109811  \\
            1.9006962888481989  688.766713261613  \\
            2.005189648552591  689.4489511390921  \\
            2.1154276726130794  690.4569620937071  \\
            2.2317261817541354  691.5429946639158  \\
            2.3544183593734545  552.9338831551395  \\
            2.4838557060784985  554.0792736536821  \\
            2.6204090466999768  555.3559375639273  \\
            2.7644695926672616  503.21114945660975  \\
            2.916450062789337  504.5858369589831  \\
            3.0767858656522016  506.07265867398746  \\
            3.24593634702017  507.6377893778052  \\
            3.4243861058147633  509.35771980195375  \\
            3.6126463824413158  511.16976067703877  \\
            3.811256523440734  513.0709515769556  \\
            4.020785526662463  515.0591463648888  \\
            4.241833671385446  483.74656793134164  \\
            4.475034238057191  485.92403217378353  \\
            4.721055322577825  488.1896035331501  \\
            4.980601750326892  490.5394482926309  \\
            5.254417095416356  492.97133056190796  \\
            5.54328581095479  574.0036429738761  \\
            5.848035476425733  576.507571293316  \\
            6.169539168618717  579.0749623570343  \\
            6.508717962905455  581.8670103320292  \\
            6.86654357202708  584.5955621587016  \\
            7.24404112995229  587.3821377892939  \\
            7.642292128781895  590.2269538100433  \\
            8.062437517113647  593.1279408981028  \\
            8.505680968743928  596.0850343353441  \\
            8.97329233107071  599.0976648666269  \\
            9.466611263077167  602.1654093213365  \\
            9.987051073318401  605.2877271987778  \\
            10.536102768906645  608.4660063059541  \\
            11.115339327095  611.6996786922505  \\
            11.726420201697199  594.6441770889844  \\
            12.371096077254009  597.8488652283105  \\
            13.051213884566325  601.1174434994012  \\
            13.76872209196403  604.4503158811273  \\
            14.52567628746955  607.8480927753326  \\
            15.324245067848398  611.310993237542  \\
            16.16671625141834  614.839983571899  \\
            17.055503432416096  618.4349498658526  \\
            17.99315289569933  622.0962238910639  \\
            18.982350911593706  625.823446409527  \\
            20.025931431784098  629.6168344102979  \\
            21.1268842082979  633.4757593498965  \\
            22.28836335884041  637.3999629853716  \\
            23.5136964030212  641.3888142125088  \\
            24.806393795359305  645.4415676133578  \\
            26.170158982378247  649.5573943793262  \\
            27.60889901260365  660.877205897938  \\
            29.12673572985976  783.4444031355042  \\
            30.728017581932694  788.1231468747096  \\
            32.41733207843099  792.8486511784748  \\
            34.19951893353395  796.3741259496724  \\
            36.07968393128038  683.5102826112723  \\
            38.06321355312053  688.5574514851622  \\
            40.155790409637426  693.6810091389086  \\
            42.36340952064813  698.871236195216  \\
            44.69239549032566  704.1280735421838  \\
            47.149420626546416  709.4480654467812  \\
            49.741524056373684  714.8300786952085  \\
            52.476131892440456  720.2547532053134  \\
            55.36107850800667  725.7514200048606  \\
            58.4046289816417  731.3012506806232  \\
            61.61550277583347  736.9024416044787  \\
            65.00289871736167  742.6535729269327  \\
            68.5765213510006  748.3165665790335  \\
            72.34660874205264  754.023028315557  \\
            76.32396180736399  759.7699966015954  \\
            80.5199752588524  765.5532957011131  \\
            84.94667024819793  771.3716775587953  \\
            89.61672880621947  777.2207369238546  \\
            94.54353017560285  783.0966473216586  \\
            99.74118914107021  788.9959000727755  \\
            105.2245964668021  794.914781531364  \\
            111.00946155696226  803.4497145060938  \\
            117.11235746154256  806.790537108148  \\
            123.55076835646473  812.744636063164  \\
            130.34313963396608  818.6996064143427  \\
            137.50893074677217  824.6523568062635  \\
            145.0686709574483  830.5935596377682  \\
            153.0440181526495  836.5260264373578  \\
            161.45782089076158  842.4403644381567  \\
            170.3341838606968  848.3416239255904  \\
            179.6985369393761  854.2027794394047  \\
            189.57770804573832  860.0624040314367  \\
            200.00000000000003  865.8660526381248  \\
        }
        ;
    \addlegendentry {GSF}
    \addplot[color={rgb,1:red,0.8889;green,0.4356;blue,0.2781}, name path={60362cd6-b92e-4304-a236-ebbbc4b17ebc}, draw opacity={1.0}, line width={2}, solid]
        table[row sep={\\}]
        {
            \\
            0.0031036826232989025  470.0713006372798  \\
            64439.57848609538  470.0713006372798  \\
        }
        ;
    \addlegendentry {RSTKF-GSF RMSE = 470.07}
    \addplot[color={rgb,1:red,0.2422;green,0.6433;blue,0.3044}, name path={0591485a-6b85-4b55-9e9e-36755cf1fbea}, draw opacity={1.0}, line width={2}, dashed]
        table[row sep={\\}]
        {
            \\
            4.241833671385446  -463.0  \\
            4.241833671385446  1763.0  \\
        }
        ;
    \addlegendentry {Min GSF RMSE = 483.75  w/ scale 4.24}
\end{axis}
\end{tikzpicture}
    \end{adjustbox}
    \vspace*{-3mm}
    \caption{Total root mean square errors vs. scale of process noise covariance $Q$ . Vertical line~(green) indicates the scale of $Q$ for which the GSF estimate~(blue) attains its minimum RMSE, which is still higher than the \textit{RSTKF-GSF} estimate~(red).}
    \label{fig:TuningBenchmark}
\end{figure}
We compared our \textit{RSTKF-GSF} with tuning the process noise covariance matrix $Q$ of the GSF tracker on the same data as in Fig.~\ref{fig:GSFProblemx} and Fig.~\ref{fig:GSFProblemRSTKFx}. Manual tuning is performed by introducing a scale parameter $c$ and using $\hat{Q} = \c \cdot Q$ as our process covariance matrix. $100$ points are sampled starting from $1$ to $200$ on a logarithmic scale to use as the scale parameter $c$. Fig.~\ref{fig:TuningBenchmark} shows the comparison of the total RMSE achieved as a function of the scale parameter of $Q$. It is observed that GSF could not beat \textit{RSTKF-GSF} with a fixed process covariance matrix $\hat{Q}$.

\section{Discussion}
\label{sec:Discussion}
%
Although there is no explicit modeling of heavy-tailed noise characteristics for process and measurement noise, the GSF is still robust to the heavy-tailed measurement noise (see Fig.~\ref{fig:Exp3}). This implicit effect comes from the marginalisation of the data association variable $\theta_k$. When the measurement is corrupted with heavy-tailed noise, the resulting value will be uninformative and will be viewed as clutter. If it's seen as clutter, then the object is misdetected. 
Thus, even though heavy-tailed measurement noise characteristics are not modeled explicitly, using a mixture representation and including the misdetection hypothesis introduces measurement outlier robustness to the GSF algorithm.

We also noted that the weight calculation involves calculating the evidence term $p(z_k^{\theta_k})$, which is intractable for the \textit{RSTKF-GSF} and is approximated. In addition, an outlier rejection method called gating can prevent the calculation of unlikely data association sequences by forming a gate around each hypothesis and rejecting all measurements outside it. The most typical form is ellipsoidal gating; it assumes additive Gaussian measurement noise and is invalid for the \textit{RSTKF-GSF} case. Non-ellipsoidal gating can be used for such methods, but that is also based on a lower bound approximation of the evidence term $p(z_k^{\theta_k})$. For more details about non-ellipsoidal outlier rejection and the lower bound of the evidence under heavy-tailed measurement noise, we refer the interested reader to \cite{agamennoni_approximate_2012}.


\section{Conclusion}
We presented an extension of the Gaussian Sum Filter, a state-of-the-art method for single object tracking using radar. Instead of a Gaussian noise model, we adopted a heavy-tailed Student's t-distribution, which accounts for the system undergoing sudden state changes. We demonstrated that our algorithm, the \textit{RSTKF-GSF}, is robust to these process outliers and does not lose track as easily. Furthermore, our simulations showed that the algorithm performs equivalently to the GSF when process outliers were absent and only presents a modest increase in computational cost.

\bibliographystyle{ieeetr}
\bibliography{references-manual}
\end{document}